\documentclass[aps,prl,twocolumn,floatfix,reprint]{revtex4-1}
\usepackage{graphicx}
\usepackage{color}
\usepackage{amssymb} 
\usepackage{amsmath} 
\usepackage{natbib}

\begin{document}

\title{Observation of the symmetry of core states of a single Fe impurity in GaAs}

\author{J. Bocquel$^1$}
\author{V.R. Kortan$^2$}
\author{R.P. Campion$^3$}
\author{B.L. Gallagher$^3$}
\author{M.E. Flatt\'e$^{1,2}$}
\author{P.M. Koenraad$^1$}

\affiliation{{$^1$\,Department of Applied Physics, Eindhoven University of Technology, P.O. Box 513, 5600 MB, Eindhoven, The Netherlands} \\
{$^2$\,Optical Science and Technology Center and Department of Physics and Astronomy, University of Iowa, Iowa City, Iowa 52242, USA}\\
{$^3$\,School of Physics and Astronomy, University of Nottingham, Nottingham NG7 2RD, United Kingdom}}

\begin{abstract}
We report the direct observation of  two mid-gap core $d$ states of differing symmetry for a single Fe atom embedded in GaAs.  These states are distinguished by the strength of their hybridization with the surrounding host electronic structure. The mid-gap state of Fe that does not hybridize via $\sigma$ bonding is strongly localized to the Fe atom, whereas the other, which does, is extended and comparable in size to other acceptor states. Tight-binding calculations of these mid-gap states agree with the spatial structure of the measured wave functions, and illustrate that such measurements can determine the degree of hybridization via $\pi$ bonding of impurity $d$~states. These single-dopant mid-gap states with strong $d$~character, which are intrinsically spin-orbit-entangled, provide an opportunity for probing and manipulating local magnetism and may be of use for high-speed electrical control of single spins.
\end{abstract}

\maketitle

The electronic localization of single dopant states within the electronic energy gap of a host semiconductor provides a model pseudo-atomic system to manipulate in an effective semiconductor ``vacuum''\cite{Koenraad2011,Steger2012}. Recent  progress in single-dopant measurement and manipulation has included optical and electronic addressing of individual spin centers\cite{Jelezko2006,Balasubramanian2009}, observation of virtual internal transitions among mid-gap states\cite{Bocquel2013}, valley-orbit coupling\cite{Zwanenburg2013}, and the effects of strain on the symmetry of the electronic wave functions\cite{Yakunin2007,Doherty2012}. An individual transition-metal dopant in a tetrahedrally-bonded semiconductor can provide access to most of these phenomena\cite{Tang2006,Yakunin2007,Chanier2012}. In addition, the potential for very large impurity spin-orbit and exchange interactions has suggested new ways to probe\cite{Tang2005} and manipulate local spins and magnetic properties using electric fields\cite{Tang2006}, strain\cite{Yakunin2007}, or a surface\cite{Strandberg2011}. For a specific single substitutional transition-metal dopant in a tetrahedrally-bonded semiconductor, the electronic structure of the mid-gap states are governed by charge-transfer energies, $d$-state filling, and the compatibility of $d$-orbital symmetry with the bonding in the surrounding host\cite{Kikoin1994,McCluskey2012}. 
In the absence of spin-orbit splitting, the $d$ states of a substitutional impurity split in the crystal field into two types of states with very different symmetry relative to the host; so-called $e$ and $t_2$ states. The $t_2$ states have the same symmetry in the crystal field as the $p$ orbitals, and hence hybridize efficiently with them along the $\sigma$ bonds connecting the impurity to its four nearest neighbors. The $e$ states, in contrast, have an incompatible symmetry with the $p$ orbitals via $\sigma$ bonding, but could hybridize through the much weaker $\pi$ bonding, or through $\sigma$ bonding and spin-orbit mixing,  to the four nearest neighbors. To date the acceptor features seen for acceptors in tetrahedrally-bonded semiconductors\cite{Zheng1994,Zheng1994E,deKort2001,Yakunin2004b,Kitchen2006,Jancu2008,Richardella2009,Bocquel2013,Muhlenberend2013} have all been associated with $t_2$ symmetry, including Zn, Mn, Co and Fe. 

Here we report the direct observation with scanning tunneling microscopy (STM) and scanning tunneling spectroscopy (STS) of $d$ states that have $e$ symmetry and hybridize with the surroundings, around a single sub-surface Fe impurity substituted for a Ga atom below the (110) surface of GaAs. The hybridization is very weak for these $e$ states compared to the previously-observed $t_2$ states, which manifests in a much more localized apparent wave function for the $e$ state than the simultaneously-observed $t_2$ state around the same Fe impurity. A theoretical description of the electronic states requires a technique that can describe the wave function on tens of thousands of atoms while preserving the local orbital symmetry in the basis; this description can be implemented in an tight-binding theory that describes the electronic structure of the host using an empirical basis\cite{Chadi1977,Tang2004,Mahani2014}, and matches the $3d$ levels of the impurity from {\it ab initio} calculations, consistent with experimental measurements. With this approach, the theoretical calculations show excellent agreement with the spatial structure of the $t_2$ states, and by ignoring $3d$-$4p$ $\pi$ hybridization between the Fe and the surrounding As atoms, provides excellent agreement for the spatial structure of the $e$ states. The penetration of  $e$ states of an Fe impurity into the surrounding GaAs, even in the absence of $pd\pi$ hybridization, also suggests that the hybridization of rare-earth dopants with a surrounding tetrahedrally-bonded host may  be observable.

\begin{figure}[t]
\centering
\hspace{-0.05cm}
\includegraphics[width=85mm]{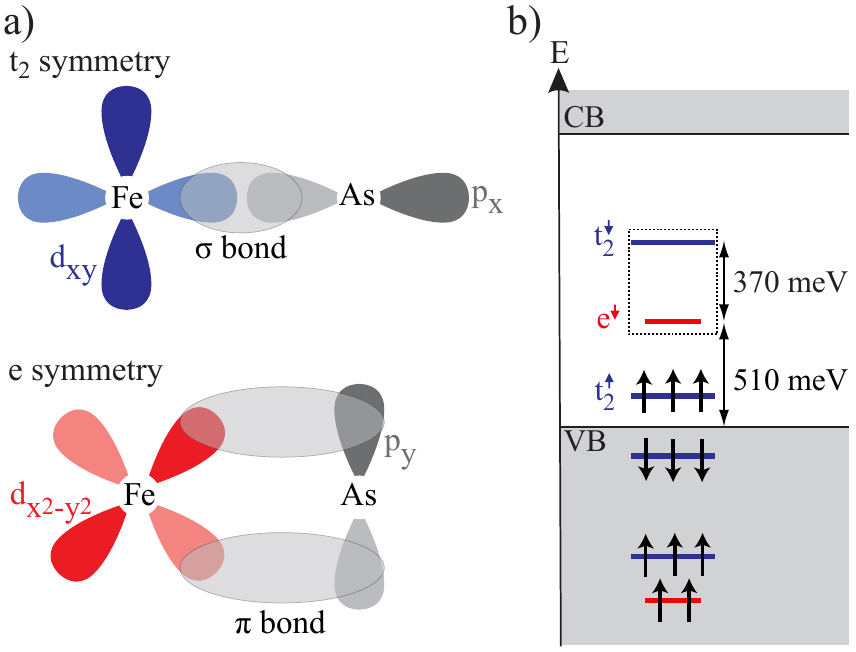}
\caption{
a) $\sigma$ and $\pi$ bonding of $3d$ transition-metal atom to one of its nearest neighbours in a tetrahedrally-bonded semiconductor along with the relative spatial orientation of the $t_2$ and $e$ orbitals. More lightly shaded orbital lobes have opposite sign amplitudes to the darker shaded lobes.
b) Energy levels and occupation of the Fe$^{3+}$ ion (with five $d$ electrons) in GaAs relative to the GaAs semiconductor bands. The $e$ impurity-like level plays an important role in the physics of charge transfer; its occupation, \textit{i.e.} its position relative to the Fermi level, determines whether the Fe impurity is in its Fe$^{3+}$ or Fe$^{2+}$ state. $e$ levels, mid-gap $t_{2}$ levels and valence-band-resonant $t_{2}$ levels are respectively indicated in red, blue and green.
}
\label{states}
\end{figure}

The spatial structure of the $t_2$ and $e$ orbitals of a $3d$ transition-metal atom in a tetrahedrally-bonded semiconductor, and the symmetry of the orbital overlaps with $p$ orbitals on the neighboring As atoms in the absence of spin-orbit interactions, are shown in Fig.~\ref{states}(a). 
The electronic configuration of the free atom is [Ar] $s^{2}d^{n}$ with $n$ electrons in the $d$~shell.  In order to replace a cation with an $s^{2}p^{1}$ electronic configuration and act as an isoelectronic impurity in the host crystal, a transition-metal atom M should release 3 electrons of which two are 4$s$ electrons and one is a 3$d$ electron. The electronic configuration becomes M$^{3+}$  (d$^{n-1}$) with $n-1$ electrons in the d-shell. The 3$d$~shell is partially filled and the 4$sp^3$ states form the outermost shell.  
Fig.~\ref{states}(b) shows the energies of the resulting features in the spectrum, with a dashed black frame around the $t_2$ and $e$ features investigated in this paper.  The different transition energies shown in this diagram were determined by optical spectroscopy \cite{Malguth2008,Dornen1993a}. 

These $t_2$ and $e$ states are observed in cross-sectional STM (X-STM) performed at 5\,K under UHV conditions (5$\times$10$^{-11}$ Torr).  Several electrochemically etched tungsten STM tips were used. The STM was operated in constant current mode on a clean and atomically flat GaAs (110) surface obtained by {\it in situ} cleavage. 
The molecular beam epitaxy grown sample contains a 100\,nm Fe-doped GaAs layer (nominal concentration of 2$\times$10$^{18}$ cm$^{-3}$) and an Fe monolayer incorporated in GaAs. The growth temperature was 480$^o$C during the entire growth procedure. The nominal layer structure consisted of GaAs substrate/100\,nm Fe:GaAs/200\,nm GaAs/Fe monolayer/500\,nm GaAs.  The two Fe-doped regions are co-doped with C atoms  (nominal concentration of 2$\times$10$^{18}$ cm$^{-3}$). These shallow acceptors greatly increase  the conductivity at the experiment's temperature of 5\,K, while having little influence on the position of the sample Fermi level (which is in the gap, close to the top of the valence band).

\begin{figure}
\centering
\hspace{-0.05cm}
\includegraphics[width=85mm]{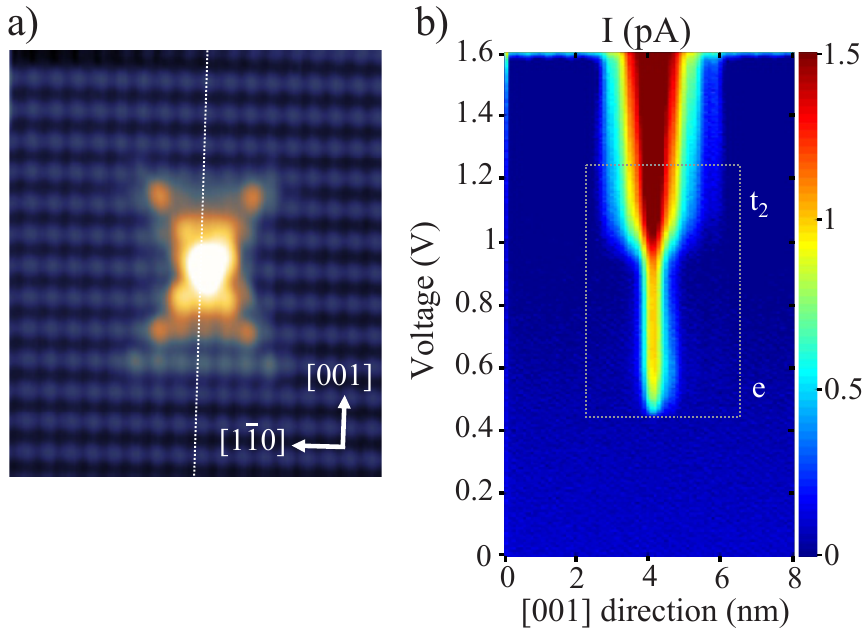}
\caption{
a) 7.5\,nm$\times$7\,nm empty states image of a single Fe impurity.
b) $dI/dV(x,V)$ cross-section taken across the Fe impurity along the [001] direction. Two peaks are resolved in the bandgap.
The onset of the conduction band is visible at a voltage of 1.55\,V.
The $dI/dV$ data taken directly on the Fe center shows two distinct peaks around  +0.55\,V and +1.0\,V, which are attributed to two states related to Fe.
}
\label{data}
\end{figure}

The empty-states topography image of  single sub-surface Fe impurity shown in Fig.~\ref{data}(a) presents a bright and anisotropic contrast. This feature shows a strong similarity with the  contrast reported for the sub-surface  [Mn$^{2+}$+h$^{+}$]  neutral acceptor state \cite{Yakunin2004b}. Both contrasts share common features like their brightness and their anisotropic shape. 
The anisotropic shape, clearly visible at low voltages, fades away at higher voltages as reported for other acceptors\cite{Yakunin2004b}.
This evolution is not completely gradual. Above V=+1.7\,V, the anisotropic shape disappears for the most part, leaving only a bright localized contrast, while a clear change in corrugation of the GaAs surface is observed. This is explained by the contribution of empty conduction band states above V=+1.7\,V, which overwhelms the smaller local density of states of the mid-gap states .

The bright electronic contrast of the Fe atom is perfectly symmetric with respect to the [001] axis and highly symmetric with respect to the [1$\overline{1}$0] axis. In the case of Mn atoms in GaAs, it has been shown that the degree of asymmetry with respect to the [1$\overline{1}$0] axis is related to the interaction between the Mn state with the asymmetric buckled surface\cite{Celebi2010}, and similar effects have been identified for Mn in InAs\cite{Marczinowski2007,Loth2008}. Consequently, the symmetry decreases as the impurities approach closer to the surface. Similar depth dependence is observed for Fe impurities.  The low Fe concentration achieved in each sample did not allow for a systematic study of Fe impurities at different depths. Nonetheless, a qualitatively similar depth dependance to Mn is observed for Fe, even if each impurity could not be unambiguously attributed to a specific depth. We estimate the Fe dopant shown in Fig.~\ref{data} to be two or three monolayers below the surface.

The Fe atoms in the first monolayers exhibit a higher degree of symmetry with respect to the [1$\overline{1}$0] direction than is seen for Mn atoms. The higher binding energy and weaker hybridization expected for Fe states with the host crystal explains this difference. In addition the surface, and the strain it produces, does not affect the wave functions of these Fe states significantly as they are more localized. Studies of the dependence of the wavefunction symmetry on acceptor binding energy indicate that the deeper the acceptor level the more symmetric it appears\cite{Celebi2008}.
The deep acceptor states of Fe are thus expected to possess a stronger impurity character than the Mn acceptor states.

In the STM  experiment, at positive bias voltages, electrons are injected in the empty states of the semiconductor sample,  that is into the conduction band and the empty energy levels associated with Fe impurities. In these conditions, the semiconductor's bands  bend upwards due to tip induced band bending (TIBB)\cite{Feenstra2002,Yakunin2004b}. The Fermi level in the bulk is in the gap, close to the top of the valence band, therefore the deep $e$ and $t_{2}$ levels, located 510\,meV and 880\,meV above the valence band edge respectively, are empty.
The majority-spin $t_{2}$ level is occupied. 
Thus the energy level occupation of Fig.~\ref{states}(b), which corresponds to the electronic configuration of the Fe$^{3+}$ isoelectronic acceptor state, applies.
This implies that electrons tunneling through the deep  $e$ and $t_{2}$ levels are responsible for the bright electronic contrast observed in the empty-states images. 
The anisotropic shape is solely attributed to the $t_{2}$ core level wave function, as the $e$ level is expected to have a much more localized contrast.
Comparing STM height profiles taken across the neutral Mn and Fe impurities shows that, in the case of Fe, the enhancement of the LDOS is more localized on the impurity itself. This is consistent with the deep nature of the Fe$^{3+}$ isoelectronic acceptors levels as well as the additional and localized tunneling channel due to the presence of an $e$ state in the bandgap for Fe. This explains why only Fe atoms a few monolayers from the surface can be resolved.

The validity of the analysis above is supported by further experimental and theoretical investigations. 
A spatially resolved I-V spectroscopy experiment at 5\,K was performed to study a single sub-surface Fe impurity.
The data acquisition was set such that the tip-sample distance was the same for every point. This is achieved by moving the tip with the feedback loop on at a voltage at which the topography is uniform across the whole image (here V=+2.5\,V). At each point, I-V curves were taken after the tip had been brought closer to the surface by 0.2-0.5\,nm with the feedback loop off. These settings are chosen to avoid any topography cross-talk in the spatial resolved I-V spectroscopy data. 

A 7.5\,nm wide $dI/dV(x,V)$ cross-section taken across the subsurface Fe impurity along the [001] direction is shown in Fig.~\ref{data}(b) for voltages between 0\,V and 1.6\,V. Each  $dI/dV$ curves has been numerically derived from the I-V curves recorded experimentally, after subtracting the offset induced by the I-V converter.
This plot shows two distinct peaks around  +0.5\,V and +1.0\,V, which are attributed to $e$ and $t_2$ acceptor states related to Fe, while the onset of the conduction band is visible at a voltage of 1.55\,V.
The fact that the $dI/dV$ signal does not drop directly  to zero at energies above these two peaks is  attributed to the tunneling from states below the Fermi energy in the tip. 

 \begin{figure}[t]
\centering
\hspace{-0.05cm}
\includegraphics[width=85mm]{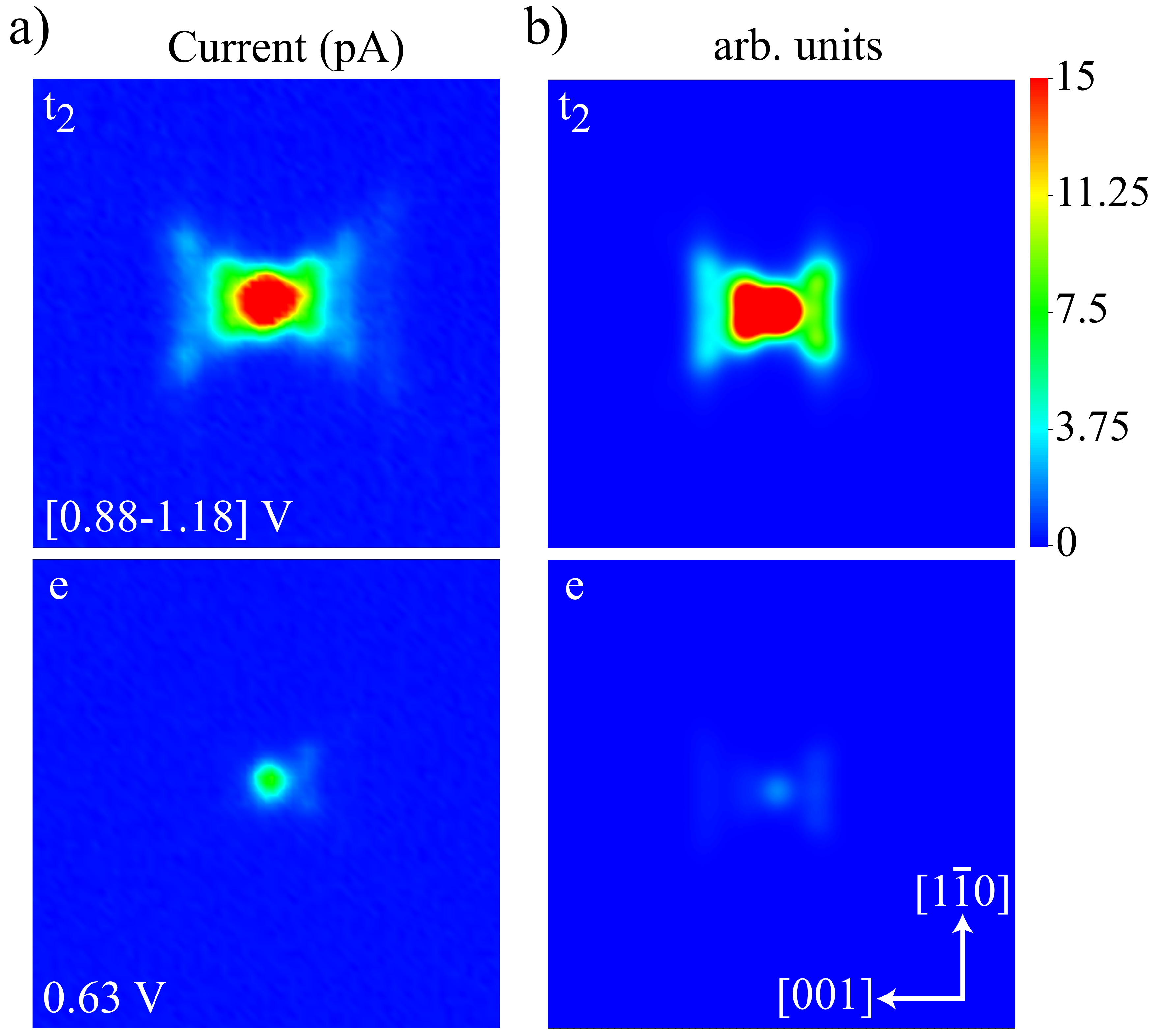}
\caption{
a) 7\,nm$\times$7\,nm experimental current ($I$) maps  integrated in a bias window [0.88-1.18]\,V (top) and taken at 0.63\,V (bottom) on a single Fe impurity at 5\,K. 
The spatial extent of these states is consistent with those expected for the deep Fe states of $t_{2}$  symmetry (higher energy) and  $e$ (lower energy).
b) 7\,nm$\times$7\,nm plots of the calculated spin averaged real space probability density of the $d$ states of $t_2$ (top) and  $e$ (bottom)  symmetry, three atomic planes away from a single Fe impurity.
}
\label{ET2_comparison}
\end{figure}

The spatial structure of these features can be most clearly seen by considering the  $I(x,y)$ maps taken at 0.63\,V and coming from the [0.88-1.18]\,V bias window, corresponding to the energy positions associated with the largest contributions to the two peaks. These are presented in  Fig.~\ref{ET2_comparison}(a) in the bottom and top panels, respectively.
The spatial extent of these two states is clearly different.
The lower-energy state is strongly localized on the Fe impurity itself. The wave function of this state is almost isotropic and extends over $\sim$ 0.75\,nm. Two small features can be seen extending in the [00$\overline{1}$] direction.
The higher localization of this state is explained by the weak hybridization of the $e$ states with states of the host crystal, due to an incompatibility of $e$-like $d$~orbitals with host $p$ orbitals [Fig.~\ref{states}(a)].
The higher energy state presents extensions into the host semiconductor in a cross-like shape.
The wave function of this state is anisotropic and extends over $\sim$ 2.5\,nm along the [001] direction and 2\,nm along the [1$\overline{1}$0] direction. 

These features are well reproduced by a tight-binding calculation, shown in Fig.~\ref{ET2_comparison}(b). We calculate the Green's functions for bulk GaAs using an $sp^3$ tight-binding description\cite{Chadi1977}. The effect of the impurity is evaluated using a Koster-Slater technique\cite{Koster1954} similar to that used to determine the acceptor state wave function for Mn in GaAs\cite{Tang2004}. Here $d$~orbitals are added to the Fe impurity site in the calculations, and the $d$-orbital energies and hopping matrix elements are introduced. The $d$-orbital energies are determined from experimental measurements of the Fe mid-gap states\cite{Malguth2008,Dornen1993a}. As the $d$~orbitals on the Fe are $3d$, whereas those included in tight-binding descriptions of GaAs\cite{Chadi1977} are $4d$, the $pd$ hopping parameters must be determined separately. We set the $pd\pi$ hopping to be zero, based on the symmetry arguments of Fig.~\ref{states}(a), so there is a single fit parameter in the theory, the $pd\sigma$ hopping, which is set based on the spatial extent of the $t_2$ state. 
The $pd\sigma$ hopping between Fe $3d$ and As $4d$ should be much smaller than that between Ga and As $4d$ states, and the optimal value we determine, $-0.0356$~eV, is two orders of magnitude smaller than that parametrized for $pd\sigma$ hopping in GaAs\cite{Jancu1998}. 
Once the extent of the $t_2$ state is set, the extent of the $e$ state is determined without adjustable parameters; the $e$ LDOS beyond the position of the Fe results from a small $t_{2}$ contribution originating from the spin-orbit interaction.   Fig.~\ref{ET2_comparison}(b) presents the calculated real space probability density of the $d$ states of $t_{2}$ (top) and $e$ (bottom) symmetry taken for a cut through the bulk GaAs crystal 3 layers away from the Fe ion.  The shapes of the calculated LDOS are in general agreement with the experimental wavefunctions and the calculated LDOS is concentrated heavily on the impurity itself.  The latter result is consistent with the experimental STM height profile taken above single Fe impurities.

\begin{figure}[t]
\centering
\hspace{-0.05cm}
\includegraphics[width=85mm]{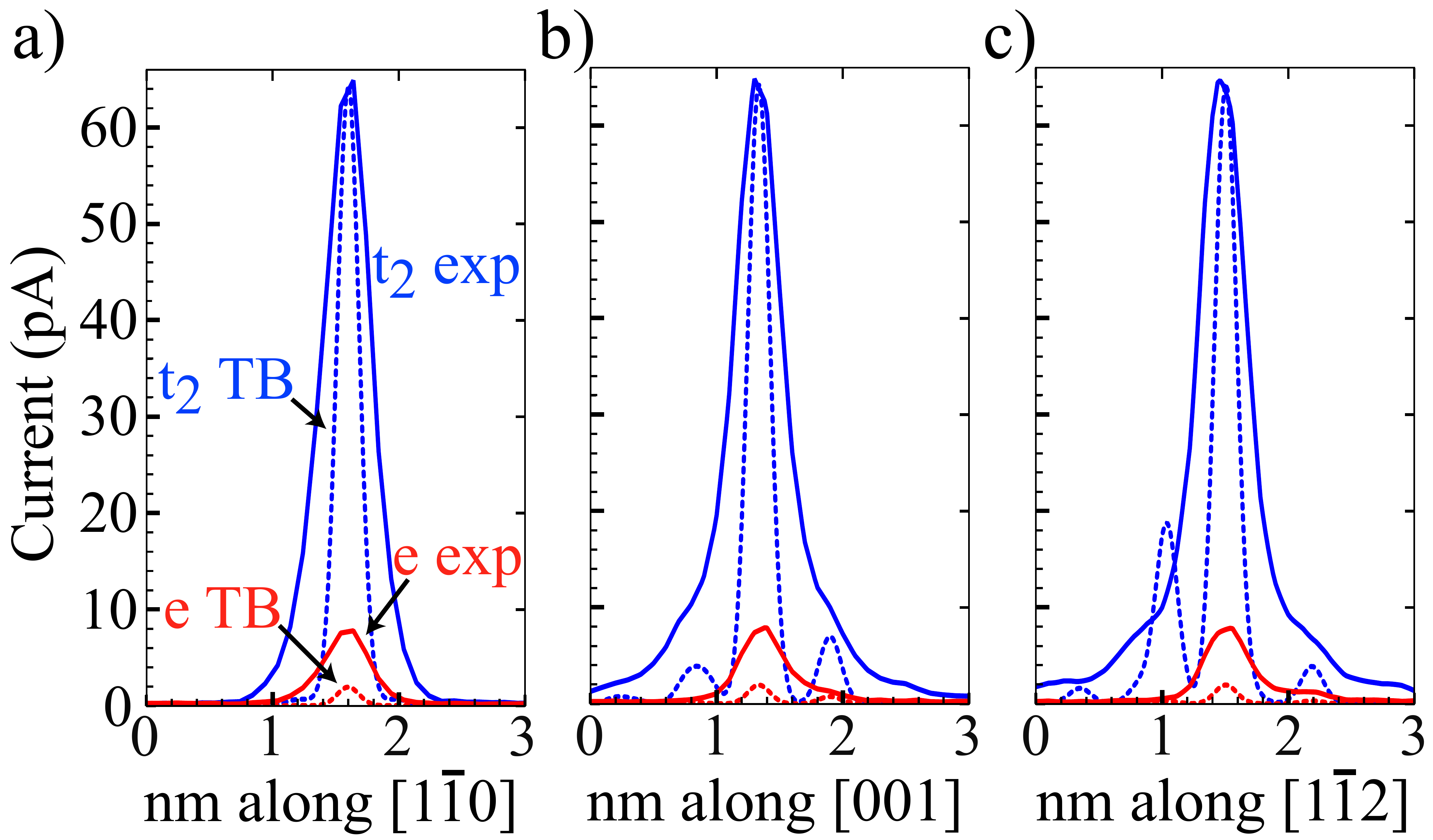}
\caption{Current measured (solid lines) and calculated (dashed lines) near a single Fe impurity at 5\,K; at 0.63\,V for the $e$ state (in red) and integrated in a bias windows [0.88-1.1.18]\,V  for the $t_2$ state (in blue) along the a) [001], b) [1$\overline{1}$0] and c [1$\overline{1}$2] directions. 
}
\label{I_profile}
\end{figure}

Agreement is also evident in Fig.~\ref{I_profile} for line cuts along three directions. Here the spatial resolution of the theoretical calculation has been set to be sharper than the experiment so that the origin of the shoulders of the $t_2$ line cuts can be more clearly seen. They originate from the amplitude of the state on neighboring atoms.

The features observed here differ greatly from those measured for Fe in the surface layer\cite{Richardella2009,Muhlenberend2013}. For Fe in the surface layer the two peaks found at 0.88\,eV and 1.5\,eV were interpreted as corresponding to splitting of the $t_{2}$ states due to symmetry-breaking at the surface. 
Their results are supported by the odd and even spatial symmetry, and similar spatial extent, of the two states appearing in the differential conductance maps at the corresponding energies\cite{Richardella2009}.
Impurity states at the surface\cite{Lee2011,Garleff2011} are known to be quite different from those even a layer below the surface, which are much less sensitive to the influence of the surface\cite{Garleff2008,Garleff2010}. 
 The two states  in our Figs.~\ref{ET2_comparison} and~\ref{I_profile}  do not exhibit  the even and odd symmetry expected for states resulting from a splitting of the $t_{2}$ state by the effect of the reconstructed surface. 
Instead, the shape and the spatial extent of these states are consistent with that expected, and calculated theoretically, if the $t_2$ states are not split.

\begin{acknowledgments}
We acknowledge support from the European Community's Seventh Framework Programme (PF7/2007-2013) SemiSpinNet and an AFOSR MURI. J.~B. and V.~R.~K. contributed equally to this work.
\end{acknowledgments}
 
 \bibliography{central-bibliography}

 \end{document}